\documentclass[aps,prl,reprint,nofootinbib,
superscriptaddress,
amsmath,amssymb,
]{revtex4-2}

\usepackage{graphicx}
\usepackage{dcolumn}
\usepackage{bm}
\usepackage{physics}
\usepackage{braket}
\usepackage{xcolor}
\usepackage{nicefrac}
\usepackage{upgreek}

\begin{document}


\title{Real-time Amplitude and Phase Estimation of AC Fields with Diamond Spins}

\author{C. T.-K. Lew}
\email{christopher.tao-kuan.lew@rmit.edu.au}
\affiliation{
School of Science, RMIT University, VIC, 3001, Australia
}

\author{S. A. Wilkinson}
\affiliation{
School of Science, RMIT University, VIC, 3001, Australia
}

\author{N. Gillespie}
\affiliation{
School of Science, RMIT University, VIC, 3001, Australia
}

\author{B. C. Gibson}
\affiliation{
School of Science, RMIT University, VIC, 3001, Australia
}

\author{D. A. Broadway}
\affiliation{
School of Science, RMIT University, VIC, 3001, Australia
}

\author{J.-P. Tetienne}
\affiliation{
School of Science, RMIT University, VIC, 3001, Australia
}

\date{\today}

\begin{abstract}
Nitrogen-vacancy centers in diamond have been shown to be capable of detecting AC magnetic fields with high sensitivity, spectral resolution, and spatial resolution. However, most studies so far have focused on the regime of time-averaged or time-correlated measurements, while little attention has been paid to the single-shot regime. Here we show that the amplitude and phase of an AC field can be retrieved from a single pair of two consecutive measurements. We demonstrate this concept by measuring a 4 MHz AC field with a per-shot amplitude and phase sensitivity of $78 \; \mathrm{nT}$ and $63 \; \mathrm{mrad}$, respectively, at a temporal resolution of $320 \; \upmu$s. We also investigate the effects and quantify the errors resulting from probe frequency detunings, as well as operating in the strong field regime. Moreover, we showcase the ability of the measurement protocol to dynamically change the probe frequency in real-time. This work advances the use of NV centers for real-time measurements of AC magnetic fields.
\end{abstract}

\maketitle







The ability to accurately detect weak AC signals is of importance for many applications, ranging from fundamental physics research \cite{Palm2022}, to nuclear magnetic resonance spectroscopy and imaging \cite{Louzon2025,Rizzato2022}, to wireless communication and localization \cite{Robertson2025,Shao2016}. Quantum sensors based on a two-level system (TLS) offer a new approach to detecting either the electric or magnetic field component of the AC signal \cite{Degen2017}. In particular, nitrogen-vacancy (NV) centers in diamond, operating as an AC magnetometer, have garnered significant interest due to their potential compact operation and ability to detecting AC fields with high sensitivity at the nano- and micro- meter length scales \cite{Barry2024,Wang2021,Loretz2013}.

Over the years, a variety of quantum AC sensing protocols have been developed, whereby the frequency, amplitude and phase information can be accurately retrieved. In instances where the phase of the AC magnetic field fluctuates measurement to measurement (i.e., stochastic and incoherent), variance-based protocols based on time-averaged measurements over multiple readouts are well suited, where over time the phase is averaged out to zero. Alternatively, quantum heterodyne (Qdyne) \cite{Louzon2025,Schmitt2017,Staudenmaier2021} and coherently averaged synchronized readout (CASR) \cite{Hermann2024,Rizzato2023} protocols have been developed for detecting coherent signals, where the phase of the AC magnetic field is fixed at the beginning of each measurement. These phase-sensitive protocols rely on measuring the frequency detuning or phase offset relative to a synchronized external clock source. Each individual measurement, typically ranging from $1-1000 \; \upmu$s, is averaged multiple times exceeding a total measurement time of seconds to obtain accurate phase estimation. In all of these examples above, the total measurement time is much greater than the duration of one individual measurement and currently little attention has been paid so far on the real-time estimation of amplitude and phase. Real-time magnetic AC sensing on the $\mathrm{\upmu s - ms}$ timescale may have potential application in probing conductive material properties via eddy current detection \cite{Chatzidrosos2019,Wickenbrock2016} and studying dynamic properties in materials enabled by magnetic AC susceptibility measurements \cite{Zhang2021,Dasika2025}.

In this work, we propose and demonstrate a real-time phase-sensitive AC detection protocol with an ensemble of NV centers in diamond. Our approach is based on a pair of sequential measurements separated by a well-defined time delay such that the orthogonal in-phase (I) and quadrature (Q) components of the AC magnetic field can be captured. For a 4 MHz test signal, a per-shot amplitude and phase sensitivity of $78 \; \mathrm{nT}$ and $63 \; \mathrm{mrad}$ was achieved, respectively, at a temporal resolution of 320\,$\upmu$s. We further systematically map out the effects and resulting errors introduced by frequency detunings and in the strong field regime (i.e., $> 5 \; \upmu$T in our experiments). Finally, we demonstrate the ability of the measurement protocol to dynamically change the probe frequency in real-time. This work expands the applicability of real-time AC sensing using quantum sensors. 


\begin{figure}[t]
    \centering
    \includegraphics[width=3.375in]{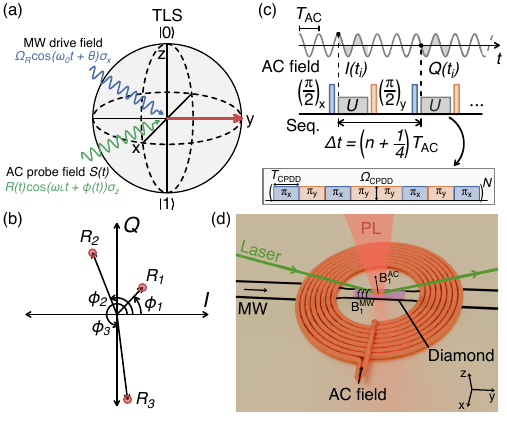}
    \caption{\textbf{Protocol for real-time magnetic field amplitude and phase estimation.} (a) Bloch sphere representation of a TLS subject to a MW driving field and an AC probe field. (b) Phasor representation of the amplitude and phase of three different measurement points, plotted in terms of its I and Q components. (c) Proposed real-time phase-sensitive AC sensing protocol. The protocol utilizes a well-defined time delay between a pair of sequential measurements. An example CPDD control sequence used in this work is shown in the inset. (d) Simplified schematic of the experimental setup. The NV diamond sample is placed on top of a CPW used to deliver the MW control pulse sequence and is surrounded by a 8-turn planar coil loop to deliver AC test signals.} 
    \label{F1}
\end{figure}

\paragraph{Sensing scheme     $-$}
We start by considering a generalized spin qubit TLS subject to a local MW driving field, as shown in Fig.~\ref{F1}(a). After bringing the qubit into a coherent superposition state in the equatorial plane by applying a $(\pi/2)_{x}$ pulse, phase is accumulated over a free evolution interval, $T_{\mathrm{evo.}}$, in response to an AC probe field, $S(t)$. Assuming a single-tone (angular) frequency, $\omega_{L}$, the Hamiltonian of the system in the laboratory frame is described by

\begin{equation}
\label{eq:eq1}
\begin{aligned}
\hat{\mathcal{H}} =& \;\hat{\mathcal{H}}_{0} + \hat{\mathcal{H}}_{MW} + \hat{\mathcal{H}}_{AC} \\
=& \;\frac{\omega_{0}}{2} \sigma_{z} + \Omega_{R} \cos{(\omega_{0}t + \theta)} \sigma_{x} \\
&+ R(t)\cos{(\omega_{L}t+\phi(t))} \sigma_{z} \; ,
\end{aligned}
\end{equation}

\noindent{where} $\omega_{0}$ is the Larmor frequency, $\Omega_{R}$ is the Rabi driving frequency with phase $\theta$, and $R(t)$ and $\phi(t)$ are the time-varying amplitude and phase of the AC magnetic field, respectively. Here, $\sigma_{x}$, $\sigma_{y}$, and $\sigma_{z}$ are the Pauli operators. The time-varying AC field can be expressed in vector representation in terms of its I and Q components (see Fig.~\ref{F1}(b))

\begin{equation}
\label{eq:eq2}
\begin{aligned}
S(t) =& \; R(t)\cos{(\omega_{L}t+\phi(t))} \\
=& \; I(t)\cos{(\omega_{L}t)}-Q(t)\sin{(\omega_{L}t)} \; ,\\
\end{aligned}
\end{equation}

\noindent{where} $I(t)=R(t)\cos{(\phi(t))}$ and $Q(t)=R(t)\sin{(\phi(t))}$. From standard trigonometric relationships, the time-varying amplitude and phase information can be trivially retrieved simultaneously (i.e., $R(t)=\sqrt{I(t)^{2}+Q(t)^{2}}$ and $\phi(t)=\mathrm{atan}{(Q(t)/I(t))}$).

The time at which we allow our qubit to accumulate phase is at the center of the proposed real-time sensing protocol, as illustrated in Fig.~\ref{F1}(c). For a coherent AC signal, a single pair of measurements is carried out to capture both amplitude and phase information at time $t_{i}$, with the second measurement in the measurement pair time delayed by $\Delta t=(n+\frac{1}{4})T_{\mathrm{AC}}$, where $n$ takes into account any measurement overhead, such as optical initialization and readout required for NV centers in diamond, and $T_{\mathrm{AC}}=2\pi/\omega_{L}=1/ f_{\mathrm{test}}$ is the period of the AC field. In contrast to previously developed protocols such as CASR that also employ time delay conditions but introduces a small frequency detuning to advance the phase of the AC signal in time at each measurement point, our protocol relies on maintaining orthogonality within each measurement pair over time.

During the sensing interval indicated by the unitary operator $U$ in Fig.~\ref{F1}(c), any arbitrary control pulse sequence may be applied to tailor the frequency response of the quantum sensor. The final $(\pi/2)_{y}$ readout pulse is chosen to be orthogonal to the first $(\pi/2)_{x}$ pulse to operate the qubit in phase-sensitive slope detection mode, where the transition probability reference point is set halfway at the point of maximum slope \cite{Degen2017,Louzon2025,Rizzato2023}. To help illustrate the proposed measurement protocol, continuous phased dynamical decoupling (CPDD) \cite{Louzon2025} control sequence shown in the inset of Fig.~\ref{F1}(c) will be applied to an ensemble of NV centers in diamond (see Supplementary Materials \cite{supp} for more details). Briefly, CPDD builds upon previous continuous \cite{Loretz2013,Wang2021,Rizzato2022,Hermann2024} and pulsed \cite{Hermann2024,Schmitt2017,Silani2023,Loretz2015,Wang2019} dynamical decoupling-based control pulse sequences by applying phase changes following the popular $XY8$ base sequence at well-defined time intervals and repeated $N$ times, suppressing both amplitude and environmental frequency noise errors. As a result, CPDD is generally more robust to experimental imperfections and as such is suitable for ensemble sensing measurements where spatial inhomogeneity of the MW driving field is unavoidable. The CPDD sequence becomes resonant with the AC field when the phase change interval and driving field strength matches the AC test frequency (i.e.,  $\omega_{L}=\pi/T_{\mathrm{CPDD}}$ and $\omega_{L}=\Omega_{\mathrm{CPDD}}$).

\paragraph{Experimental details     $-$}
Measurements were performed on a 100 $\upmu$m thick NV-doped diamond sample with the top and bottom surfaces orientated along the $\braket{111}$ direction. Laser pulses ($\approx 20 \; \upmu$s) are used to optically spin polarize the NVs to the $m_{s} = 0$ ground state and read out the final spin state population. A $B_{0}\approx 6.1$ mT bias magnetic field is applied along the $\braket{111}$ direction to lift the degeneracy between the $m_{s} = \pm1$ states of the NV electron spin, forming an effective TLS by considering only the $m_{s} =0 \rightarrow m_{s} =-1$ spin transition. 

The diamond sample is placed on top of a coplanar waveguide (CPW) printed circuit board (PCB) to efficiently drive resonant spin transitions, as shown in Fig.~\ref{F1}(d). The final $(\pi/2)_{y}$ readout pulse is phased cycled between 0 and $180^{\circ}$ to cancel out low frequency noise arising from charge state and laser intensity fluctuations \cite{Hart2021}. A 8-turn planar coil loop terminated to a 50~$\Omega$ load \cite{Herb2020} surrounding the diamond sample is used to deliver AC test signals at 4 MHz with amplitude $R_{\mathrm{test}} = 2\;\upmu$T, unless otherwise specified. Green laser is illuminated from the side of the diamond and the resulting photoluminescence (PL) is collected from the top with a photodetector. Full experimental details can be found in the Supplemental Materials \cite{supp}.

\begin{figure}[t]
    \centering
    \includegraphics[width=3.375in]{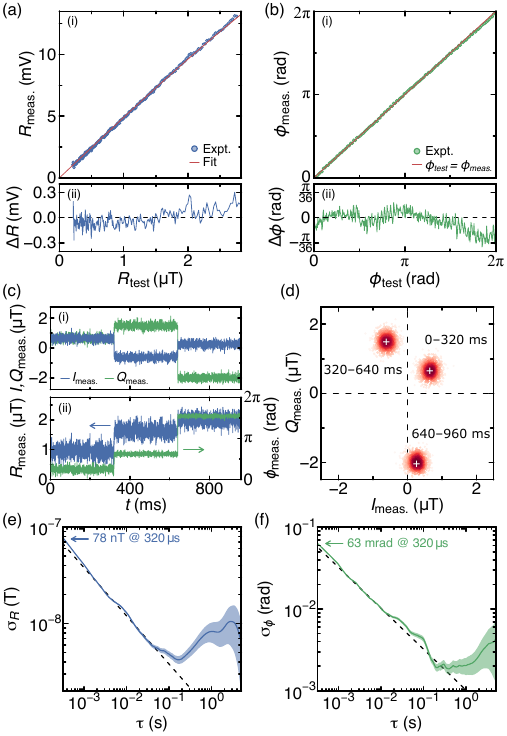}
    \caption{\textbf{Demonstration of real-time AC sensing.} (a)(i) Measured PL response as a function of input AC magnetic field strength. The red solid line is expected behavior from theory following Eqs.~\ref{eq:eq3} and \ref{eq:eq4}, with the slope of the fit equal $4.85$ mV/$\upmu$T. The residual is plotted in (ii). (b) Measured phase response with the phase of the test field swept from [0, $2\pi$]. The red solid line represents the ideal case where $\phi_{\mathrm{test}} = \phi_{\mathrm{meas.}}$ and the residual is plotted in (ii). (c) Time trace of the measured (i) in-phase, quadrature, (ii) amplitude and phase component of a time-varying AC test field, with amplitude and phase changes applied every $320$ ms. The corresponding IQ diagram is plotted in (d) and the white crosses indicate expected location based on $R_{\mathrm{test}}$ and $\phi_{\mathrm{test}}$. For each set point (white cross), 1000 data points were measured (i.e., $320 \; \upmu \mathrm{s}$ per data point over 320 ms) and represented as a density map with dark (light) red indicating high (low) sample density. (e) Amplitude and (f) phase Allan deviation calculated from a 10~s time trace. Shaded areas represent the uncertainty.} 
    \label{F2}
\end{figure}

\paragraph{Results     $-$}
We first calibrate the measured PL (in mV) in response to an AC test signal using the CPDD-XY8 control sequence repeated $N=14$ times (i.e., $T_{\mathrm{evo.}}=8T_{\mathrm{CPDD}}N=14 \; \upmu$s) by experimentally increasing the AC amplitude in Fig.~\ref{F2}(a)(i). The magnetic field strength is related to the power via a conversion factor, $\alpha = R_{\mathrm{test}}/\sqrt{P}_{\mathrm{AC}}$ \cite{Vallabhapurapu2021,Mariani2022}, which was experimentally predetermined to be approximately equal to $670 \; \upmu \mathrm{T/\sqrt{W}}$ for the 8-turn planar coil loop (see Supplementary Materials \cite{supp} for more information). 

The solution to the Hamiltonian in Eq.~\ref{eq:eq1} expressed in terms of a normalized PL contrast from phase cycling the final $(\pi/2)_{y}$ pulse for the I and Q component can be described by \cite{Louzon2025}

\begin{subequations}
\begin{align}
I(t) =& \: A(T_{\mathrm{evo.}}) \sin{(\kappa \gamma_{NV} R_{\mathrm{test}} T_{\mathrm{evo.}} \cos{(\phi(t))})} \notag \\ 
\approx& \: A(T_{\mathrm{evo.}})\xi \cos{(\phi(t))} , \label{eq:eq3}  \\
Q(t) =& \: A(T_{\mathrm{evo.}}) \sin{(\kappa \gamma_{NV} R_{\mathrm{test}} T_{\mathrm{evo.}} \cos{(\phi(t) + \pi/2)})} \notag \\ 
\approx& \: A(T_{\mathrm{evo.}}) \xi \cos{(\phi(t) + \pi/2)} , \label{eq:eq4}
\end{align}
\end{subequations}

\noindent{where} $A(T_{\mathrm{evo.}})$ is the signal contrast, $\xi=\kappa\gamma_{NV}R_{\mathrm{test}} T_{\mathrm{evo.}}$ with $\kappa$ being the attenuation factor and equal to $1/2$ for CPDD \cite{Louzon2025}, and $\gamma_{NV}$ is the NV center electron spin gyromagnetic ratio. Note the last set of (linear) approximations are valid only in the weak magnetic field regime (i.e., $\xi \cos{(\phi(t))}, \xi \cos{(\phi(t) + \pi/2)} \ll 1$). 

We fit the experimental data in Fig.~\ref{F2}(a)(i) to relate the measured PL response to the AC magnetic field strength. The experimental data agrees well with theory from Eqs.~\ref{eq:eq3} and ~\ref{eq:eq4} (red solid line) in the weak field limit, further highlighted by the residual plotted in Fig.~\ref{F2}(a)(ii). Additionally, the slope of the fitted red solid line is determined to be equal to $4.85$ mV/$\upmu$T, which will be used for the remainder of this work. We then verified the phase response in Fig.~\ref{F2}(b) by sweeping the phase of the test signal from [0, $2\pi$], revealing a near one-to-one phase correspondence with minimal residuals (see Fig.~\ref{F2}(b)(ii)). 

We now showcase the real-time phase-sensitive measurement protocol in response to a test signal with its amplitude, $R_{\mathrm{test}}= \{ 0.9,1.6,2.1 \} \; \upmu $T, and phase, $\phi_{\mathrm{test}} = \{\nicefrac{\pi}{4}, \nicefrac{5\pi}{8},\nicefrac{37 \pi}{24} \} \; \mathrm{rad}$, varied every $320$ ms. The measured I and Q components, as well as the corresponding calculated amplitude and phase are plotted in Fig.~\ref{F2}(c)(i) and (ii), respectively. Accounting for measurement overhead, the temporal resolution of this measurement is 320 $\upmu$s. An alternative method to represent Fig.~\ref{F2}(c)(i) is in terms of an IQ diagram shown in Fig.~\ref{F2}(d), with each data point corresponding to one IQ measurement every 320 $\upmu$s and the three white crosses indicate expected values, $R_{\mathrm{test}}$ and $\phi_{\mathrm{test}}$. The majority of measured data points lie close to the expected values, demonstrating excellent real-time amplitude and phase estimation of a dynamically changing AC signal.

To quantitatively benchmark the current sensitivity, we calculate the amplitude and phase Allan deviation over 10~s in Figs.~\ref{F2}(e) and (f), respectively. An approximate $\tau^{-1/2}$ behavior (black dotted lines) expected for white Gaussian noise is observed up to $\approx100$ ms, before deviating at longer timescales, indicating an increase in colored noise possibly due to temperature and background magnetic field drift \cite{Wolf2015}. A per-shot amplitude and phase sensitivity of $78 \; \mathrm{nT}$ and $63 \; \mathrm{mrad}$ was extracted, respectively, at a temporal resolution of 320 $\upmu$s. Moreover, a minimum in the Allan deviation of $4.2 \: \mathrm{nT}$ and $1.9 \: \mathrm{mrad}$ is observed at $\tau = 160 $ ms and $\tau = 390 $ ms in Fig.~\ref{F2}(e) and (f), respectively. The extracted normalized amplitude and phase sensitivity are approximately $1.7 \: \mathrm{nT/\sqrt{Hz}}$ and $1.2 \: \mathrm{mrad/\sqrt{Hz}}$, respectively. These values are comparable to previous demonstration with Qdyne at GHz frequencies \cite{Staudenmaier2021} but at least $1-3$ orders of magnitude less sensitive than state-of-the-art implementation at MHz frequencies enabled by employing magnetic flux concentrators \cite{Silani2023}. This suggests that significant enhancement in the sensitivity is possible, which is beyond the scope of our proof-of-principle demonstration. 

\begin{figure}[t]
    \centering
    \includegraphics[width=3.375in]{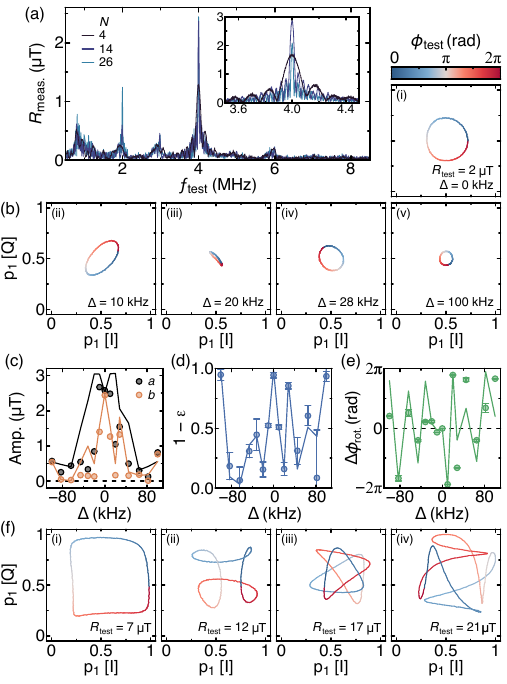}
    \caption{\textbf{Errors arising from frequency detunings and large amplitudes.} (a) Measured amplitude response as a function of AC test frequency swept in $10$ kHz steps, with the CPDD sequence resonantly tuned to $f_{\mathrm{probe}}=4$ MHz for different values of $N$. A separate set of measurements scanned over a smaller frequency range in $1$ kHz steps is shown in the inset. (b) IQ diagrams at selective frequency detunings, with the test signal phase swept from [0, $2\pi$]. (c) Extracted major axis, semi-major axis and the (d) calculated ellipticity as a function of frequency detuning. (e) Measured phase rotation as a function of frequency detuning. Solid lines are expected behavior calculated using an idealized model. (f) Selective IQ diagrams in the high field non-linear regime with the test signal phase swept from [0, $2\pi$].} 
    \label{F3}
\end{figure}

Having demonstrated real-time amplitude and phase sensing, we now investigate errors arising from operating our ensemble spin system beyond normal conditions by first exploring the effects of frequency detunings. By setting the CPDD resonance condition to $f_{\mathrm{probe}} = 4$ MHz and sweeping the AC test frequency for different values of $N$ repetitions, we observe non-zero amplitude contribution at specific frequencies, as shown in Fig.~\ref{F3}(a). A sharp peak in the measured amplitude response is observed and reaches a maximum when $f_{\mathrm{test}}=f_{\mathrm{probe}}$. The corresponding full width half maximum (FWHM) linewidth narrows with increasing $N$ following $2 f_{\mathrm{probe}}/8N$, as expected for the corresponding sinc filter function in the frequency domain associated with the CPDD control sequence \cite{Louzon2025}. Beyond the filter function bandwidth, additional spurious harmonics are observed at multiples of $f_{\mathrm{probe}}$. These undesired harmonics may be possibly suppressed by introducing randomized global phase changes for each $N$ repetition \cite{Loretz2015,Wang2019}, in addition to MW pulse shaping \cite{Fuchs2009}.

Additional insights about the effects of frequency detunings (i.e., $\Delta =f_{\mathrm{test}} - f_{\mathrm{probe}}$) can be gathered by performing full phase sweeps from [0, $2 \pi$], as shown in the selective IQ diagrams in Fig.~\ref{F3}(b). For reference, an ideal IQ diagram with zero detuning is shown in Fig.~\ref{F3}(b)(i) and the full set of detuning data can be found in the Supplementary Materials \cite{supp}. Qualitatively, at certain detunings the amplitude in the IQ diagram is compressed in one direction, forming an ellipsoid rather than an ideal circle. In addition, a phase rotation is observed, accompanied by a reversal in the direction of rotation. These peculiar features are quantified and summarized in Figs.~\ref{F3}(c)$-$(e). Here, the ellipticity is defined as the difference in diameter between the major and semi-minor axis, normalized to the major axis (i.e., $\epsilon = (a-b)/a$). The quantity $1-\epsilon$ plotted in Fig.~\ref{F3}(d) indicates a perfect circle when equal to 1 and an ellipsoid when equal to 0. In Fig.~\ref{F3}(e), the sign of $\Delta \phi_{\mathrm{rot.}}$ indicates the  direction of rotation (i.e., $+ = \mathrm{counterclockwise}$, $- = \mathrm{clockwise}$). 

Using an idealized model indicated by the solid lines in Figs.~\ref{F3}(c)$-$(e), the majority of features experimentally observed are reproduced quite well by theory. These features are generally a result of frequency detuning-induced phase offset relative to when I and Q measurements are sampled (see Supplemental Materials \cite{supp} for more details). Further work is required to understand the remaining discrepancies, which may be explained by extending the idealized model to include experimental imperfections and decoherence processes.

For completeness, we also explore measurement errors introduced when operating in the high field non-linear regime in Fig.~\ref{F3}(f). Significant distortion and departure from a perfect circle is observed at large amplitudes and is explained in terms of phase wrapping as the spin state population must fall between 0 and 1 \cite{Degen2017}. This phase wrapping behavior can be well explained by the idealized model (see Supplemental Materials \cite{supp}). Phase wrapping may be partially alleviated by applying adaptive sensing techniques \cite{Degen2017} or alternative protocols \cite{Zhang2022,Nusran2013} to address the limited dynamic range associated with quantum sensing utilizing slope detection.

The asymmetry in the IQ diagrams may be partially explained by Rabi detuning introduced by hyperfine interaction with the nitrogen nuclear spin \cite{Louzon2025,Staudenmaier2021}. These effects may be suppressed by operating at higher magnetic fields to spin polarize the nitrogen nuclear spin \cite{Wood2016,Busaite2020}. However, other forms of broadening and experimental imperfections may also play a significant role in the observed departure from the idealized model. 

\begin{figure}[t]
    \centering
    \includegraphics[width=3.375in]{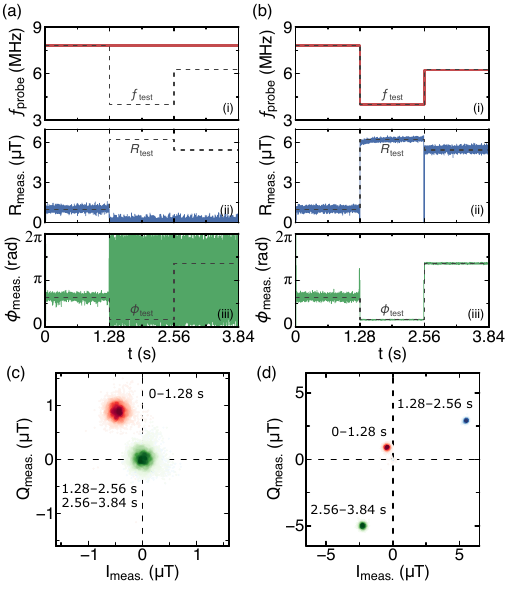}
    \caption{\textbf{Demonstration of real-time frequency tracking.} Measured (ii) amplitude and (iii) phase response of a time-varying AC test signal with frequency changes applied every 1.28 s (a) without and (b) with frequency tracking enabled by dynamically adjusting the probe frequency in (i). Each measurement was taken every $320 \; \upmu$s, totaling 4000 data points per frequency change. Dotted gray lines indicate expected set values. Corresponding IQ diagrams (c) without and (d) with frequency tracking. Each color indicates a different probe frequency and the sample density is represented by the shade of each color (i.e., dark (bright) = high (low) sample density).} 
    \label{F4}
\end{figure}

Finally, we extend the applicability of the measurement protocol by implementing dynamic frequency tracking to perform amplitude and phase estimation of a time-varying AC test signal with its frequency, amplitude, and phase varied simultaneously. For this demonstration, the test signal frequency, $f_{\mathrm{test}} = \{7.8125,4,6.25\} \; \mathrm{MHz}$, amplitude, $R_{\mathrm{test}} = \{1,6.25,5.45\} \; \mathrm{ \upmu T}$, and phase,  $\phi_{\mathrm{test}} = \{\nicefrac{23 \pi}{36}, \nicefrac{11 \pi}{72},\nicefrac{49 \pi}{36} \} \; \mathrm{rad}$, are varied every 1.28 s, as indicated by the gray dotted lines in Figs.~\ref{F4}(a)$-$(b). In the case where $f_{\mathrm{probe}}$ is held constant at 4 MHz set by the CPDD parameters and the time delay $\Delta t$ in Fig.~\ref{F4}(a), the measured amplitude and phase deviates from expected set values as the test frequency is varied. In other words, the I and Q components approaches 0, as shown in Fig.~\ref{F4}(c). In constrast, $f_{\mathrm{probe}}$ is dynamically adjusted by changing $\Delta t$ and the CPDD parameters to match $f_{\mathrm{test}}$ in Figs.~\ref{F4}(b) and (d). The measured amplitude and phase are in excellent agreement with $R_{\mathrm{test}}$ and $\phi_{\mathrm{test}}$ except near when $f_{\mathrm{test}}$ is changed (attributed to imperfect synchronization), demonstrating real-time frequency tracking at a temporal resolution of $320 \; \upmu$s. For this measurement, the number of repetitions in the CPDD sequence, $N = \{27,14,22\}$, is also dynamically adjusted to keep the CPDD filter function FWHM linewidth relatively fixed at $\approx 71$ kHz. However, this is not necessary and the filter function linewidth can be arbitrarily set and dynamically adjusted to increase temporal resolution at the expense of sensitivity or vice versa.  


\paragraph{Discussion     $-$}
The temporal resolution of the measurement protocol can be easily reduced by at least a factor of 4 for the same CPDD-XY8, $N=14$ sequence as currently the measurement overhead dominated by the laser pulse duration is the limiting factor. With sufficiently high laser power intensity, a pulse duration of only $\approx 2 \; \upmu$s is required, including the necessary post-laser wait time governed by the photodynamics of the NV center \cite{Tetienne2012,Dreau2011}. Additional reduction in measurement overhead time may be possible by simultaneously addressing multiple spin transitions using orthogonal control pulse sequences \cite{Chen2024}. Temporal resolution may be traded off with improved sensitivity by extending the number of $N$ repetitions up to the coherence time of the NV center. Beyond that, engineering the diamond host material to extend the NV center coherence time \cite{Barry2024} may further improve the sensitivity. Additional improvement may be achieved by enhancing the photon collection efficiency \cite{Barry2024,LeSage2012} and considering MW control pulse sequences with corresponding tailored-made filter functions \cite{Cangemi2025,Biercuk2011}. 

\paragraph{Conclusion     $-$}
We have proposed and demonstrated a real-time AC sensing protocol to measure the amplitude and phase of a time-varying oscillating magnetic field based on consecutive time-offset measurements of a spin ensemble. A per-shot amplitude and phase sensitivity of of $78 \; \mathrm{nT}$ and $63 \; \mathrm{mrad}$ was achieved, respectively, at a temporal resolution of $320 \; \upmu$s. We further analyzed errors introduced by operating outside intended normal conditions by probing test signal frequencies slightly detuned from the resonance condition set by the control sequence, as well as in the strong field regime. As a final demonstration, we show that the sensing protocol can be dynamically tuned to maintain accurate amplitude and phase estimation despite frequency changes in the test AC field. 

\begin{acknowledgments}
This work was supported by the Australian Government under the Advanced Strategic Capabilities Accelerator through its Emerging and Disruptive Technologies (EDT) Program (in partnership with Diamond Defence Pty Ltd. and Phasor Innovation Pty Ltd.). D. A. B. acknowledges support from the Australian Research Council through grant DE230100192. S. A. W. would like to thank Genko Genov for helpful advice. N. G. acknowledges support by the Defence Science Institute, an initiative of the State Government of Victoria.
\end{acknowledgments}


\bibliography{apssamp}

\end{document}